\documentclass[aps,prd,reprint,superscriptaddress,nofootinbib]{revtex4-1}
\usepackage{amsmath}
\usepackage{amssymb}
\usepackage{graphicx}
\usepackage[caption=false]{subfig}
\usepackage{enumitem}
\usepackage{color}
\usepackage{mathtools}
\usepackage[percent]{overpic}
\usepackage{bm}
\usepackage{rotating}
\usepackage{cancel}
\usepackage[hidelinks]{hyperref}

\newcommand{\ket}[1]{{\left| {#1} \right>}}
\newcommand{\bra}[1]{{\left< {#1} \right|}}

\newcommand{\ii}{\mathrm{i}}
\newcommand{\trf}{\text{Tr}_{\hat{\phi}}}

\newcommand{\ac}{\lambda}

\DeclareMathOperator{\Realpart}{Re}

\begin{document}

\title{Degenerate detectors are unable to harvest spacelike entanglement}

\author{Alejandro Pozas-Kerstjens}
\affiliation{ICFO-Institut de Ciencies Fotoniques, The Barcelona Institute of Science and Technology, 08860 Castelldefels (Barcelona), Spain}

\author{Jorma Louko}
\affiliation{School of Mathematical Sciences, University of Nottingham, Nottingham NG7 2RD, United Kingdom}

\author{Eduardo Mart\'{i}n-Mart\'{i}nez}
\affiliation{Institute for Quantum Computing, University of Waterloo, Waterloo, Ontario N2L 3G1, Canada}
\affiliation{Department of Applied Mathematics, University of Waterloo, Waterloo, Ontario N2L 3G1, Canada}
\affiliation{Perimeter Institute for Theoretical Physics, 31 Caroline St. N., Waterloo, Ontario N2L 2Y5, Canada}


\begin{abstract}
We show, under a very general set of assumptions, that pairs of identical particle detectors in spacelike separation, such as atomic probes, can only harvest entanglement from the vacuum state of a quantum field when they have a nonzero energy gap. Furthermore, we show that degenerate probes are strongly challenged to become entangled through their interaction through scalar and electromagnetic fields even in full light-contact. We relate these results to previous literature on remote entanglement generation and entanglement harvesting, giving insight into the energy gap's protective role against local noise, which prevents the detectors from getting entangled through the interaction with the field.
\end{abstract}

\maketitle


\section{Introduction}

It is well-known that the ground state of a quantum field contains entanglement between different regions of spacetime. This is so even if the regions are spacelike separated \cite{Summers1985,Summers1987}. Moreover, this entanglement can be extracted (or \textit{harvested}) into pairs of particle detectors through local interactions of each detector with the field (again, even in spacelike separation), leading to the entanglement of initially uncorrelated detectors \cite{Valentini1991,Reznik2003,Reznik2005} even for arbitrary spatial separation and smooth switching profiles \cite{Pozas-Kerstjens2015}.

This phenomenon, known as \textit{entanglement harvesting}, is very sensitive to the properties of the spacetime background (e.g., its geometry \cite{Steeg2009} or its topology \cite{Martin-Martinez2016a}).

Entanglement harvesting has been proposed as a means to build sustainable sources of entanglement (via \textit{entanglement farming} protocols \cite{Martin-Martinez2013a}), and has been proven to be very sensitive to the state of motion of the detectors and the boundary conditions on the field on which it is performed. This has led to proposals of applications in metrology such as range finding \cite{Salton2015} or as a very sensitive means to detect vibrational motion \cite{Brown2014}. 

Entanglement harvesting has been proven to be substantially independent of the particular particle detector model employed: there are no notable qualitative differences between simplified Unruh-DeWitt models in its different variants. Namely, it was shown in Ref. \cite{Pozas-Kerstjens2016} that an Unruh-DeWitt detector coupled to the amplitude or to the momentum of a scalar field yields qualitatively similar results to those of a fully featured hydrogenlike atom coupled to the electromagnetic field. Harmonic oscillator detectors have also been shown to display the same qualitative behaviour when they harvest entanglement from the quantum field \cite{Brown2013}. Along these lines, entanglement harvesting is not a fragile phenomenon: it has been proven robust against uncertainties in the synchronization and spatial configuration of the particle detectors \cite{Barry}. The variety of situations in which the phenomenon of entanglement harvesting has been found relevant has motivated works analyzing the experimental feasibility of implementing timelike and spacelike entanglement harvesting protocols in both atomic and superconducting systems \cite{Olson2011,Olson2012,Sabin2012}.

Entanglement harvesting is  affected by local noise. For example, a sudden switching of the detector-field interaction (which locally excites the detectors) is inefficient for harvesting spacelike entanglement since the local noise overshadows the correlations harvested from the field. In contrast, if the interaction is switched on adiabatically, it has been shown that it is possible to harvest entanglement with arbitrarily distant spacelike separated detectors \cite{Pozas-Kerstjens2015}. To harvest spacelike entanglement from arbitrarily long distances, the detectors' energy gaps (the energy difference between ground and first excited state) have to be increased proportionally to the separation of the detectors to shield them from local excitations that would overwhelm the harvesting of correlations (see Ref. \cite{EduSingleAuthor} for a thorough study).

It has been observed that temperature also prevents entanglement from being harvested \cite{Brown2013single}, particularly for spacelike separation between the detectors. This can be understood as caused by the decay of quantum correlations in a quantum field with temperature.

Remarkably, and in contrast to this, it was shown by Braun \cite{Braun2002,Braun2005} that, even with zero energy gap, spin-1/2 systems in timelike separation could entangle through their interaction with thermal baths and quantum fields in thermal states. This mechanism was initially proposed as a means of creating entanglement between distant parties \cite{Braun2002}, but a closer examination of the problem revealed that the more interesting phenomenon of spacelike entanglement harvesting---in which not even indirect communication through the field is possible and none of the detectors can know of the existence of the other---was not possible in the cases studied in Ref. \cite{Braun2005}. These results raise the question of what is special in the regimes analyzed in Refs. \cite{Braun2002,Braun2005} that prevents spacelike entanglement harvesting. In principle, and with no additional data, one could have thought of three possible suspects for the lack of spacelike entanglement harvesting in the setups in Refs. \cite{Braun2002,Braun2005}: 1) the use of thermal backgrounds as opposed to the vacuum state of the field, 2) the particular switching functions utilized (recall that switching can strongly influence the ability to harvest entanglement \cite{Reznik2005,Pozas-Kerstjens2015}) or 3) the fact that \cite{Braun2002,Braun2005} only analyze degenerate two-level systems (with zero gap between ground and excited states).

In this paper we address this question and show that the lack of spacelike entanglement harvesting is not due to the thermal background or to the nature of the switching. The culprit is the gapless nature of the detectors. We prove that it is impossible for a pair of identical inertial gapless detectors to harvest any amount of entanglement from spacelike separated regions even in the vacuum state of a scalar field in flat spacetime, and argue that the proof should carry over to the case of entanglement harvesting with hydrogenlike atoms from the electromagnetic field \cite{Pozas-Kerstjens2016}.

After an introduction to the formalism of entanglement harvesting and the notation to be used throughout the paper in Sec. \ref{sec:formalism}, we divide the proof in two parts: in Sec. \ref{sec:nooverlap} we prove that when the time intervals of interaction of each individual detector with the field do not overlap, gapless detectors cannot harvest any entanglement \textit{at all}, regardless of their specific spatial shape, their relative separation (not only spacelike, but also timelike or lightlike) or the total amount of time of interaction with the field, and then in Sec. \ref{sec:overlap} we give the proof that spacelike entanglement harvesting is not possible in the case when the periods of interaction have nonzero overlap, which requires the extra assumption of the shapes being spherically symmetric. In Sec. \ref{sec:EM} we extend the results in Secs. \ref{sec:nooverlap} and \ref{sec:overlap} to detectors interacting with an electromagnetic field through a realistic dipole-type light-matter interaction. In Sec. \ref{sec:deltas} we also show that very short and strong `delta-like' switching functions cannot harvest entanglement at all regardless of energy gaps, regime of separation or smearing of the detectors. Finally, in Sec. \ref{sec:summary} we conclude by providing the physical interpretation of the results: as was already noted in Ref. \cite{Pozas-Kerstjens2015}, the energy gap shields from local excitations of the detectors and its absence allows for any local noise to overcome the nonlocal excitations produced by the vacuum fluctuations.

\section{Unruh-DeWitt dynamics and entanglement harvesting}\label{sec:formalism}

In a typical scenario of entanglement harvesting \cite{Valentini1991,Reznik2003}, two localized quantum systems interact with the vacuum state of a field. We model the interaction between an individual inertial smeared detector and a massless scalar field in an ($n+1$)-dimensional flat spacetime with the Unruh-DeWitt (UDW) particle detector model \cite{DeWittBook}. This model captures the fundamental features of the light-matter interaction in scenarios where angular momentum exchange does not play a fundamental role \cite{Martin-Martinez2013,Alhambra2014,Pozas-Kerstjens2016}. More relevant to our case, the UDW model has been explicitly proven to yield qualitatively identical results in entanglement harvesting to those with fully featured hydrogenoid atoms interacting with the electromagnetic field (in particular, see Ref. \cite{Pozas-Kerstjens2016} for this last claim).  For technical reasons, we assume throughout $n\ge2$. The case $n=1$ would require additional input for handling the well-known infrared divergences of a massless field in two spacetime dimensions. We will make some explicit comments about the 1+1 dimensional case when we discuss some of our results. 

The UDW interaction Hamiltonian is given by
\begin{equation}
    \hat{H}(t)=\!\sum_{\nu}\ac_\nu \mathcal{X}_\nu\left(t\right)\!\!\int\!\text{d}^n\bm x\,S_\nu\!\left(\bm x-\bm x_\nu\right)\hat{\mu}_\nu(t)\hat{\phi}(t,\bm x).
    \label{Hamiltonian}
\end{equation}

In this expression, the label $\nu\in\{\text{A},\text{B}\}$ identifies the detector and $\ac_\nu$ is the coupling strength of detector $\nu$ to the scalar field $\hat{\phi}(t,\bm x)$. The field can be written as a sum of plane-wave modes as
\begin{equation}
    \hat{\phi}(t,\bm{x})=\int\frac{\text{d}^n\bm{k}}{\sqrt{(2\pi)^n2|\bm k|}}
\left[\hat{a}_{\bm{k}}e^{\ii \mathsf{k}\cdot \mathsf{x}}+\hat{a}^\dagger_{\bm{k}}e^{-\ii \mathsf{k}\cdot \mathsf{x}}\right],
\end{equation}
where $\hat{a}_{\bm{k}}$ and $\hat{a}_{\bm{k}}^\dagger$ are bosonic annihilation and creation operators of a field mode with momentum $\bm k$, and \mbox{$\mathsf{k}\cdot\mathsf{x}=-|\bm k|t+\bm k\cdot\bm x$}. $\hat{\mu}_\nu(t)$ is the monopole moment of detector $\nu$, given by
\begin{equation}
    \hat{\mu}_\nu(t)=e^{\ii\Omega_\nu t}\hat{\sigma}^{+}_\nu+e^{-\ii\Omega_\nu t}\hat{\sigma}^{-}_\nu
    \label{monopole}
\end{equation}
($\hat{\sigma}^{+}$ and $\hat{\sigma}^{-}$ are the usual SU(2) ladder operators). Here $\Omega_\nu$ is the energy gap between the two levels of detector~$\nu$. $\mathcal{X}_\nu(t)$ is the \textit{switching function} that controls the duration and strength of the interaction. $S_\nu(\bm x)$ is the \textit{smearing function} of the detectors that can be associated to their spatial extension and shape (e.g., for a hydrogenoid atom it is connected to the ground and excited state wavefuctions \cite{Pozas-Kerstjens2016}).

As usual in entanglement harvesting scenarios, the detectors, initially completely uncorrelated and in their ground state, couple to the field, and after the coupling (controlled by the switching function), they end up in a final state given by
\begin{equation}
    \hat{\rho}_\textsc{ab}=\trf\left(\hat{U}\ket{\psi_0}\bra{\psi_0}\hat{U}^\dagger\right),
\end{equation}
where $\trf$ denotes the partial trace with respect to the field degrees of freedom. Here
\begin{equation}
    \hat{U}=\mathcal{T}\exp\left(-\ii\int_{-\infty}^{\infty}\!\!\!\!\text{d}t\,\hat{H}(t)\right)
\end{equation} 
is the time evolution operator and the initial state of the detectors-field system is taken to be
\begin{equation}
    \ket{\psi_0}=\ket{g_\textsc{a}}\otimes\ket{g_\textsc{b}}\otimes|0_{\hat{\phi}}\rangle.
\end{equation}

We consider detectors that have identical energy gaps and identical spatial shapes, so that 
\mbox{$\Omega_\textsc{a}=\Omega_\textsc{b}\equiv\Omega$}
and 
\mbox{$S_\textsc{a}=S_\textsc{b}\equiv S$}. 
We also take the coupling strengths to be identical, so that 
\mbox{$\ac_\textsc{a}=\ac_\textsc{b}\equiv\ac$},
and the switching functions to be identical up to a time shift, 
so that \mbox{$\mathcal{X}_\nu(t) \equiv \mathcal{X}(t - t_\nu)$},
where $t_\nu$ is the time at which the interaction of detector $\nu$ and the field begins.

The detectors' state $\hat{\rho}_\textsc{ab}$ after the interaction is a two-qubit X-state \cite{Reznik2003,Pozas-Kerstjens2015}. 
We quantify the entanglement in this state with the negativity 
(a faithful entanglement measure for a system of two qubits \cite{Vidal2002}). 
To the first nontrivial perturbative order in the coupling strength, 
the negativity takes the simple form \cite{Reznik2005,Pozas-Kerstjens2015}
\begin{equation}
    \mathcal{N}^{(2)}=\text{max}(0,|\mathcal{M}|-\mathcal{L})+\mathcal{O}(\ac^4).
    \label{negativity}
\end{equation}
Note that throughout this paper we are using the notation in Ref. \cite{Pozas-Kerstjens2015}. The functions $\mathcal{L}$ and $\mathcal{M}$ are
\begin{align}
\mathcal{L}=&\ac^2\int_{-\infty}^{\infty}\!\!\!\!\text{d}t_1\int_{-\infty}^{\infty}\!\!\!\!\text{d}t_2\,\mathcal{X}(t_1)\mathcal{X}(t_2)e^{\ii\Omega(t_1-t_2)}\notag\\
&\times\int\text{d}^n\bm{x}_1\int\text{d}^n\bm{x}_2\,S(\bm{x}_1)S(\bm{x}_2)W_n(t_2,\bm{x}_2,t_1,\bm{x}_1)\notag\\
=&\ac^2\int\text{d}^n\bm k \frac{|\tilde{S}(\bm{k})|^2}{2|\bm{k}|}\int_{-\infty}^{\infty}\!\!\!\!\text{d}t_1\,\mathcal{X}(t_1)e^{\ii(|\bm{k}|+\Omega)t_1}\notag\\
&\times\int_{-\infty}^{\infty}\!\!\!\!\text{d}t_2\,\mathcal{X}(t_2)e^{-\ii(|\bm{k}|+\Omega)t_2}\notag\\
=&\ac^2\int\text{d}^n\bm k \frac{|\tilde{S}(\bm{k})|^2}{2|\bm{k}|}\left|\int_{-\infty}^{\infty}\!\!\!\!\text{d}t\,\mathcal{X}(t)e^{\ii(|\bm{k}|+\Omega)t}\right|^2,
\label{L}
\end{align}
\begin{align}
\mathcal{M}=&-\ac^2\int_{-\infty}^{\infty}\!\!\!\!\text{d}t_1\int_{-\infty}^{t_1}\!\!\!\!\text{d}t_2\int\text{d}^n\bm{x}_1\int\text{d}^n\bm{x}_2\notag\\
&\times S(\bm{x}_1-\bm{x}_\textsc{a})S(\bm{x}_2-\bm{x}_\textsc{b})\notag\\
&\times e^{\ii\Omega(t_1+t_2)}W_n(t_1,\bm{x}_1,t_2,\bm{x}_2)\notag\\
&\times\left[\mathcal{X}\left(t_1\!-\!t_\textsc{a}\right)\mathcal{X}\left(t_2\!-\!t_\textsc{b}\right)\!+\!\mathcal{X}\left(t_1\!-\!t_\textsc{b}\right)\mathcal{X}\left(t_2\!-\!t_\textsc{a}\right)\right]\notag\\
=&-\ac^2\int\text{d}^n\bm k\,\frac{|\tilde{S}(\bm{k})|^2}{2|\bm{k}|} e^{\ii\bm{k}\cdot(\bm{x}_\textsc{a}-\bm{x}_\textsc{b})} \notag\\
&\int_{-\infty}^{\infty}\!\!\!\!\text{d}t_1\int_{-\infty}^{t_1}\!\!\!\!\text{d}t_2\,e^{-\ii(|\bm k|-\Omega)t_1} e^{\ii(|\bm{k}|+\Omega)t_2}\notag\\
&\times\!\left[\mathcal{X}\left(t_1\!-\!t_\textsc{a}\right)\mathcal{X}\left(t_2\!-\!t_\textsc{b}\right)\!+\!\mathcal{X}\left(t_1\!-\!t_\textsc{b}\right)\mathcal{X}\left(t_2\!-\!t_\textsc{a}\right)\right],
\label{M}
\end{align}
the Wightman function of the free scalar field in $n$ spatial dimensions is given by
\begin{equation}
W_n(t,\bm x,t',\bm x')=\langle0_{\hat{\phi}}|\hat{\phi}(t,\bm x)\hat{\phi}(t',\bm x')|0_{\hat{\phi}}\rangle,
\end{equation}
and the Fourier transform of the smearing function is
\begin{equation}
    \tilde{S}(\bm k)=\frac{1}{\sqrt{(2\pi)^n}}\int\text{d}^n\bm x\,S(\bm x)e^{\ii\bm k\cdot\bm x}. 
    \label{Fourier}
\end{equation}

 We have used the time translation invariance of $W_n(t,\bm x,t',\bm x')$ 
to write \eqref{L} in a way that 
makes explicit that $\mathcal L$ is independent of the beginning of the interaction with the field $t_\nu$.

It is already discussed in Refs. \cite{Reznik2003,Reznik2005}, and with our notation in Refs. \cite{Pozas-Kerstjens2015,Pozas-Kerstjens2016}, that the term $\mathcal{L}$ corresponds to local excitations of each detector, while $\mathcal{M}$ accounts for correlations between both detectors. Therefore, Eq. \eqref{negativity} has an intuitive physical meaning: for two detectors to harvest entanglement from the field (i.e., for the negativity of the joint state $\hat{\rho}_\textsc{ab}$ to be nonzero after interacting with the field) the correlation term $\mathcal{M}$ must overcome the local noise $\mathcal{L}$. 

Our objective is to prove that identical zero-gap detectors cannot harvest entanglement from spacelike separated regions of the field.

From now on we consider gapless detectors, \mbox{$\Omega=0$}, 
so that the monopole moment \eqref{monopole} becomes time independent. 
We also take the switching function 
$\mathcal{X}$ to have compact support, writing 
\begin{equation}
\mathcal{X}\left(t\right)=\begin{cases}
\chi(t) &   \text{for}\,\,0\leq t\leq T\\
0           &   \text{otherwise}
\end{cases},
\label{switching}
\end{equation}
where $T>0$ is the duration of each detector's 
interaction with the field. 
We emphasize that the times~$t_\nu$, at which the interaction 
of each detector with the field begins, remain arbitrary. 
These initial times have dropped out of ${\mathcal L}$ \eqref{L} 
but they appear in ${\mathcal M}$ \eqref{M}. 
Similarly, we emphasize that the spatial 
points~$\bm x_\nu$, at which the detectors are centered,
have dropped out of ${\mathcal L}$ \eqref{L} 
but they appear in ${\mathcal M}$ \eqref{M}.



\section{Non overlapping switchings}\label{sec:nooverlap}

When the switching functions' domains do not overlap, the time integrals in the nonlocal term \eqref{M} greatly simplify. There are two summands in this term, which require separate study.

In the first summand the integrand is nonzero for $t_1\in [t_\textsc{a},\,t_\textsc{a}+T]$ and $t_2\in [t_\textsc{b},\,t_1\leq \text{min}(t_\textsc{a},t_\textsc{b})+T]$. Without loss of generality, let us assume that detector B is switched on after detector A has been switched off (i.e., $t_\textsc{b}>t_\textsc{a}+T$). In this case, because of the nested nature of the integrals, the region of integration over $t_2$ is limited by the support of $\mathcal{X}(t_1-t_\textsc{a})$. Since detector B is switched on after detector A is switched off, the region of integration over $t_2$ lies out of the support of $\mathcal{X}(t_1-t_\textsc{a})$, and therefore the integral evaluates to 0 regardless of the specific shape of $\chi(t_2)$.

In the second summand, in contrast, the integrand is supported in $t_1\in [t_\textsc{b},\,t_\textsc{b}+T]$ and \mbox{$t_2\in [t_\textsc{a},\,t_1\leq \text{min}(t_\textsc{a},t_\textsc{b})+T]$}. Now, in the case that detector B is switched on after detector A has been switched off, the effective region of integration over $t_2$ after taking into account the supports of $\mathcal{X}(t_1-t_\textsc{b})$ and $\mathcal{X}(t_2-t_\textsc{a})$ is $[t_\textsc{a},\,t_\textsc{a}+T]$. This means that we can denest the two integrals,

\begin{align}
    \int_{-\infty}^{\infty}&\!\!\!\!\text{d}t_1\int_{-\infty}^{t_1}\!\!\!\!\text{d}t_2\,e^{-\ii|\bm k|(t_1-t_2)} \mathcal{X}\left(t_1\!-\!t_\textsc{b}\right)\mathcal{X}\left(t_2\!-\!t_\textsc{a}\right)\\
    \notag &=
    \int_{-\infty}^{\infty}\!\!\!\!\text{d}t_1\int_{-\infty}^{\infty}\!\!\!\!\text{d}t_2\,e^{-\ii|\bm k|(t_1-t_2)} \mathcal{X}\left(t_1\!-\!t_\textsc{b}\right)\mathcal{X}\left(t_2\!-\!t_\textsc{a}\right),
\end{align}
where the equality follows because all the values of $t_2$ in the support of $\mathcal{X}\left(t_2\!-\!t_\textsc{a}\right)$ are strictly smaller than the smallest value of $t_1$ in the support of $\mathcal{X}\left(t_1\!-\!t_\textsc{b}\right)$.

Now, using the fact that the modulus of an integral is upper bounded by the integral of the modulus of the integrand, i.e.,
\begin{equation}
    \left|\int\text{d}x\,f(x)\right|\leq\int\text{d}x\left|f(x)\right|,
    \label{absintegralineq}
\end{equation}
we see that
\begin{align}
|\mathcal{M}|=&\ac^2\Bigg|\int\text{d}^n\bm k\,\frac{|\tilde{S}(\bm{k})|^2}{2|\bm{k}|}e^{\ii\bm{k}\cdot(\bm{x}_\textsc{a}-\bm{x}_\textsc{b})}\notag\\
&\times\!\int_{-\infty}^{\infty}\!\!\!\!\!\!\text{d}t_1\!\!\int_{-\infty}^{\infty}\!\!\!\!\!\!\text{d}t_2\,e^{\ii|\bm{k}|(t_1-t_2)}\mathcal{X}\!\left(t_1-t_\textsc{b}\right)\!\mathcal{X}\!\left(t_2-t_\textsc{a}\right)\!\Bigg|\notag\\
\leq&\ac^2\int\text{d}^n\bm k \frac{|\tilde{S}(\bm{k})|^2}{2|\bm{k}|}\notag\\
&\times\Bigg|\int_{-\infty}^{\infty}\!\!\!\!\text{d}t\int_{-\infty}^{\infty}\!\!\!\!\text{d}t'\,e^{\ii|\bm{k}|(t-t')}\mathcal{X}(t)\mathcal{X}(t')\Bigg|=\mathcal{L},\label{nlproof}
\end{align}
where the changes of variables \mbox{$t=t_1-t_\textsc{b}$} and $t'=t_2-t_\textsc{a}$ have been performed.

This yields the following conclusion: when there is no overlap between the time intervals of individual interactions with the field, the nonlocal, entangling term is always upper-bounded by the local one and therefore \mbox{$\mathcal{N}^{(2)}=\text{max}\left(0,|\mathcal{M}|-\mathcal{L}\right)= 0$} for any---compactly supported or not---smearing function of the (recall gapless) detectors and any compactly supported, nonoverlapping switchings.

This means that gapless inertial comoving detectors with the same switching functions are unable to harvest any entanglement regardless of their relative positioning (spacelike, timelike or lightlike) from even arbitrarily close regions if they are switched on at different times with no overlap between the time intervals in which each individual detector interacts with the field.

We would like to stress that this is the case even for gapless detectors which are in regions that can be connected by light. This is true even if the smearing functions overlap (which means having effectively zero distance between the detectors).

Although this proof assumed that the switchings were the same for both detectors, numerical evidence for a generality of compactly supported switching functions suggests that the detectors are unable to harvest entanglement also in the case of switchings of different duration $T_\textsc{a}\neq T_\textsc{b}$. We highlight that this is true for detectors in timelike, spacelike or even lightlike separation.

Finally, notice that this proof carries over to the case of 1+1 dimensions if we add an infrared cutoff. Even with an infrared cutoff, the identity \eqref{absintegralineq} still holds in the same way as in \eqref{nlproof}, so the inability of gapless detectors to harvest entanglement applies also to this case. 

\section{Overlapping switchings}\label{sec:overlap}

We now explore the case when the time intervals of interaction overlap, either partially or totally. For this scenario, numerical evidence shows that entanglement harvesting is possible in general for timelike and lightlike separations, so we focus on the harvesting of entanglement from spacelike separated regions and ask the following question: can two gapless detectors harvest entanglement from the field vacuum while they remain spacelike separated?

To talk properly about spacelike separation, we consider detectors with arbitrary compactly supported smearings. Concretely, detectors A and B have finite characteristic lengths of $R_\textsc{a}$ and $R_\textsc{b}$ respectively. In analogy with Eq. \eqref{switching}, the smearing functions of the detectors are given by
\begin{equation}
S_\nu\left(\bm x\right)=\begin{cases}
s_\nu(\bm x) &  \text{for}\,\,\left|\bm x\right|\leq\frac{1}{2}R_\nu\\
0           &   \text{otherwise}
\end{cases}.
\end{equation}

For the following proof, we furthermore assume that the shapes of the detectors are spherically symmetric, which amounts to saying that their Fourier transform as given by Eq. \eqref{Fourier} only depends on the norm of the Fourier variable $\bm k$. Explicitly, writing \eqref{L} and \eqref{M} in spherical coordinates
\begin{align}
    \mathcal{L}=&\ac^2\int_0^\infty\!\!\!\!\!\!\text{d}|\bm k|\!\int\!\text{d}\Omega_{n-1} |\bm k|^{n-2}\frac{|\tilde{S}(|\bm{k}|)|^2}{2}\!\left|\int_{-\infty}^{\infty}\!\!\!\!\!\!\text{d}t\,\mathcal{X}(t)e^{\ii|\bm{k}|t}\right|^2,\label{Langles}\\
    \mathcal{M}=&-\ac^2\int_0^\infty\!\!\!\!\text{d}|\bm k|\int\text{d}\Omega_{n-1} |\bm k|^{n-2}\,\frac{|\tilde{S}(|\bm{k}|)|^2}{2} e^{\ii\bm{k}\cdot(\bm{x}_\textsc{a}-\bm{x}_\textsc{b})} \notag\\
&\int_{-\infty}^{\infty}\!\!\!\!\text{d}t_1\int_{-\infty}^{t_1}\!\!\!\!\text{d}t_2\,e^{-\ii|\bm k|(t_1-t_2)} \notag\\
&\times\!\left[\mathcal{X}\left(t_1\!-\!t_\textsc{a}\right)\mathcal{X}\left(t_2\!-\!t_\textsc{b}\right)\!+\!\mathcal{X}\left(t_1\!-\!t_\textsc{b}\right)\mathcal{X}\left(t_2\!-\!t_\textsc{a}\right)\right].
\label{Mangles}
\end{align}

The spherical symmetry of the smearing allows us to perform the integration over the angular variables that appear in Eqs. \eqref{Langles} and \eqref{Mangles}. 
On the one hand, the integrals in the local term \eqref{L} straightforwardly evaluate to the surface of the $(n-1)$-sphere, while on the other hand the integrals in the nonlocal term \eqref{M} are slightly less straightforward and are computed explicitly in appendices \ref{app:nonlocalangleintegral} and \ref{app:time}. The resulting expressions for $\mathcal{L}$ and $\mathcal{M}$ are
\begin{align}
\mathcal{L}=&\ac^2\int_0^\infty\!\!\!\!\text{d}|\bm k||\bm k|^{n-2}\,|\tilde{S}(|\bm{k}|)|^2\frac{\pi^{\frac{n}{2}}}{\Gamma(n/2)}\text{Re}\left[\mathcal{T}_0(|\bm k|,T)\right],\label{Lmoments}\\
\mathcal{M}=&-\ac^2\int_0^\infty\!\!\!\!\text{d}|\bm k||\bm k|^{n-2} |\tilde{S}(|\bm{k}|)|^2\frac{\pi^{\frac{n}{2}}}{\Gamma(n/2)}\notag\\
&\times{}_0F_1\left(\frac{n}{2};-\frac{|\bm k|^2|\bm x_\textsc{a}-\bm x_\textsc{b}|^2}{4}\right) \mathcal{T}_{\Delta t}(|\bm k|,T),
\label{Mmoments}
\end{align}
where ${}_0F_1\left(a;z\right)$ is the confluent hypergeometric limit function \cite{NIST:DLMF},
\begin{align}
    \mathcal{T}_{\Delta t}&(|\bm k|,T) =\int_{-\infty}^{\infty}\!\!\!\!\text{d}t_1\int_{-\infty}^{\infty}\!\!\!\!\text{d}t_2\,\theta(t_1-t_2)e^{-\ii|\bm k|(t_1-t_2)}\notag\\
    &\times\left[\mathcal{X}\left(t_1\right)\mathcal{X}
    \left(t_2- \Delta t\right)
    +\mathcal{X}\left(t_1-\Delta t\right)
    \mathcal{X}\left(t_2\right)\right], 
    \label{TtB}
\end{align}
and $\Delta t = t_\textsc{b} - t_\textsc{a}$.


A crucial observation to prove that gapless detectors with overlapping interaction time intervals cannot harvest spacelike entanglement is that only the real part of the function $\mathcal{T}_{\Delta t}$ contributes to the evaluation of $\mathcal{M}$ when the detectors are spacelike separated. To see this, we return to the expression of $\mathcal{M}$ in terms of the Wightman function in Eq. \eqref{M}, which for gapless detectors is
\begin{align}
\mathcal{M}=&-\ac^2\int_{-\infty}^{\infty}\!\!\!\!\text{d}t_1\int_{-\infty}^{t_1}\!\!\!\!\text{d}t_2\int\text{d}^n\bm{x}_1\int\text{d}^n\bm{x}_2\notag\\
&\times S(\bm{x}_1-\bm{x}_\textsc{a})S(\bm{x}_2-\bm{x}_\textsc{b})W_n(t_1,\bm{x}_1,t_2,\bm{x}_2)\notag\\
&\times\left[\mathcal{X}\left(t_1\!-\!t_\textsc{a}\right)\mathcal{X}\left(t_2\!-\!t_\textsc{b}\right)\!+\!\mathcal{X}\left(t_1\!-\!t_\textsc{b}\right)\mathcal{X}\left(t_2\!-\!t_\textsc{a}\right)\right].
\label{MWightman}
\end{align}

Given that the smearing and switching functions are real, the only element that can make $\mathcal{M}$ complex is the Wightman function $W_n$. Remarkably, the imaginary part of the Wightman function $W_n(t,\bm x,t',\bm x')$ is proportional to (the expectation value of) the commutator of the field at the points $(t,\bm x)$ and $(t^\prime,\bm x^\prime)$. Namely (see e.g, Eq. (23) in Ref. \cite{EduSingleAuthor}),
\begin{equation}
    \langle 0_{\hat{\phi}}|[\hat{\phi}(t,\bm x),\hat{\phi}(t',\bm x')]| 0_{\hat{\phi}}\rangle=2\ii\, \text{Im}\Big[W_n(t,\bm x,t',\bm x')\Big].
    \label{Wightmancomm}
\end{equation}

The commutator between field observables (and in particular, the field commutator) is only supported inside their respective light cones (this property is known as \textit{micro-causality}). Therefore, for spacelike separated regions the imaginary part of the Wightman function as given by Eq. \eqref{Wightmancomm} vanishes and the nonlocal term described by Eq. \eqref{MWightman} is real. This means, from \eqref{MWightman}, that $\mathcal{M}$ is real.

Armed with this information about $\mathcal{M}$, we look at it in the form \eqref{Mmoments}. Since the hypergeometric functions in Eq. \eqref{Mmoments} are real and $\mathcal{M}$ itself is real, we conclude that only the real part of $\mathcal{T}_{\Delta t}(|\bm k|,T)$ contributes to $\mathcal{M}$. This allows us to replace $\mathcal{T}_{\Delta t}(|\bm k|,T)$ by $\text{Re}\left[\mathcal{T}_{\Delta t}(|\bm k|,T)\right]$ for any switching and radially symmetric smearing under the condition that the detectors are spacelike separated.

Continuing with the proof, 
we show in appendix \ref{app:TtB} that
\begin{align}
    \text{Re}\left[\mathcal{T}_{\Delta t}(|\bm k|,T)\right]&=2\pi|\tilde{\mathcal{X}}(|\bm k|)|^2\cos(|\bm k|\Delta t)\notag\\
    &=\text{Re}\left[\mathcal{T}_0(|\bm k|,T)\right]\cos(|\bm k|\Delta t).
    \label{TtBsimple}
\end{align}

As the confluent hypergeometric limit function satisfies (see 10.14.4 and 10.16.9 in Ref. \cite{NIST:DLMF})
\begin{equation}
    \left|{}_0F_1\left(\alpha;-x^2\right)\right|\leq 1, 
\end{equation}
we obtain
\begin{align}
    |\mathcal{M}|=&\ac^2\Bigg|\int_0^\infty\!\!\!\!\text{d}|\bm k||\bm k|^{n-2}|\tilde{S}(|\bm k|)|^2\frac{2\pi^{\frac{n}{2}+1}}{\Gamma(n/2)}\notag\\
    &\times{}_0F_1\left(\frac{n}{2};-\frac{|\bm k|^2|\bm x_\textsc{a}-\bm x_\textsc{b}|^2}{4}\right)|\tilde{\mathcal{X}}(|\bm k|)|^2\cos(|\bm k|\Delta t)\Bigg|\notag\\
    &\leq \ac^2\int_0^\infty\!\!\!\!\text{d}|\bm k||\bm k|^{n-2}|\tilde{S}(|\bm k|)|^2\frac{2\pi^{\frac{n}{2}+1}}{\Gamma(n/2)}|\tilde{\mathcal{X}}(|\bm k|)|^2=\mathcal{L}.
\end{align}

This implies that $\mathcal{N}^{(2)}=0$ for gapless, spacelike separated spherically symmetric detectors for any zero, or nonzero overlap between the time intervals of interaction of each detector with the field. Hence, combining the results of this section with those of Sec. \ref{sec:nooverlap}, we see that gapless detectors with finite, spherically symmetric smearings interacting for a finite time with the field can never harvest entanglement from spacelike separated regions, independently of the specific way of interacting with the field or their shape. This, of course, includes as a particular case the use of pointlike detectors, which is the case that is used most often in the literature.

\section{A very relevant nonspherically symmetric case: The realistic light-matter interaction}\label{sec:EM}

In this section we consider the realistic case of the light-matter interaction. Namely, the interaction of an atomic electron in a hydrogenlike atom with the vacuum state of an electromagnetic field through a dipolar coupling. Our study becomes particularly relevant for  transitions between orbitals of the same quantum number $n$, which have zero energy gap. In the simplified case of pointlike atoms, there was numerical evidence that gapless detectors do not allow for entanglement harvesting in spacelike separated regions \cite{Braun2005}.  

Beyond that simplification, the general study of atom-light interactions for arbitrary finite energy gaps was reported in Ref. \cite{Pozas-Kerstjens2016}, where the fully featured shape of the atomic wave functions was taken into account. In particular, it was shown in Ref. \cite{Pozas-Kerstjens2016} that entanglement harvesting from both electromagnetic and scalar fields exhibits the same qualitative features despite the difference in the setups. We now focus on the case of two fully featured hydrogenlike atoms when an energy degenerate transition is used to harvest entanglement from the vacuum state of the electromagnetic field.

For a pair of identical atoms, the negativity takes a similar form as in the scalar case. Namely, the negativity acquired after interaction is given by Eq. \eqref{negativity} where the local $\mathcal{L}$ and nonlocal $\mathcal{M}$ terms become now (see. Eqs. (31) and (32) in Ref. \cite{Pozas-Kerstjens2016})
\begin{align}
\mathcal{L}^\textsc{em}=&\,e^2\int_{-\infty}^{\infty}\!\!\!\!\text{d}t_1\int_{-\infty}^{\infty}\!\!\!\!\text{d}t_2\, \mathcal{X}(t_1)\mathcal{X}(t_2)\notag\\
&\times\!\!\int\text{d}^3\bm x_1\int\text{d}^3\bm x_2
\,{\bm{S}^*}^{\text{\textbf{t}}}(\bm x_2)\text{\bf W}(t_2,\bm x_2,t_1,\bm x_1)\bm S(\bm x_1),\label{LEMWightman}\\
\mathcal{M}^\textsc{em}=&-e^2\int_{-\infty}^{\infty}\!\!\!\!\text{d}t_1\int_{-\infty}^{t_1}\!\!\!\!\text{d}t_2\int\text{d}^3\bm x_1\int\text{d}^3\bm x_2\notag\\
&\times\Big[\mathcal{X}(t_1-t_\textsc{a})\mathcal{X}(t_2-t_\textsc{b})\notag\\
&\quad\times{\bm{S}_\textsc{a}}^{\!\!\text{\textbf{t}}}(\bm x_1-\bm x_\textsc{a})\text{\bf W}(t_1,\bm x_1,t_2,\bm x_2)\bm S_\textsc{b}(\bm x_2-\bm x_\textsc{b})\notag\\
&+\mathcal{X}(t_1-t_\textsc{b})\mathcal{X}(t_2-t_\textsc{a})\notag\\
&\quad\times{\bm{S}_\textsc{b}}^{\!\!\text{\textbf{t}}}(\bm x_1-\bm x_\textsc{b})\text{\bf W}(t_1,\bm x_1,t_2,\bm x_2)\bm S_\textsc{a}(\bm x_2-\bm x_\textsc{a})\Big],
\label{MEMWightman}
\end{align}
where $e$ is the electron charge, the matrix $\text{\bf W}(t_1,\bm x_1,t_2,\bm x_2)$ is the is the Wightman tensor of the electric field operator $\bm{\hat{E}}$ whose components are given by
\begin{equation}
\left[\text{\bf W}\right]_{ij}=W_{ij}(t,\bm x,t',\bm x')=\langle 0_{\bm{\hat{E}}}|\hat{E}_i(t,\bm x)\hat{E}_j(t',\bm x')|0_{\bm{\hat{E}}}\rangle,
\end{equation}
and the vectors ${\bm{S}_\nu}^{\!\!\text{\textbf{t}}}$ and ${\bm{S}_\nu^*}^{\text{\textbf{t}}}$ are respectively the transpose and Hermitian conjugate of the vector ${\bm{S}_\nu}$ (the spatial smearing vector) which relates to the ground and excited wave functions by
\begin{equation}
    \bm S_\nu(\bm x)=\psi^*_{e_\nu}(\bm x)\bm x\psi_{g_\nu}(\bm x)
\end{equation}
(note that this smearing vector is called $\bm F_\nu(\bm x)$ in \cite{Pozas-Kerstjens2016}).

In the case of atomic switching functions that do not overlap, the reasoning used in Sec. \ref{sec:nooverlap} applies: the first summand of Eq. \eqref{MEMWightman} evaluates to 0 and in the second summand the integrals in time denest, making $|\mathcal{M}^\textsc{em}|$ upper-bounded by $\mathcal{L}^\textsc{em}$, regardless of the smearing vectors being compactly supported or not. This means that non-simultaneously interacting hydrogenlike atoms cannot harvest any entanglement from the vacuum  at all through transitions of zero energy.

When there is some overlap between the intervals of interaction of each individual atom with the field, the arguments used in Sec. \ref{sec:overlap} would also apply for hypothetical compactly supported  atoms: in this case, and since the electric field also satisfies micro-causality (the electric field commutator is zero for spacelike separated events), $\mathcal{M}^\textsc{em}$ would also be real for spacelike separations between the compactly supported atoms. Then, without assuming spherical symmetry of the smearing functions, the hypergeometric function in Eq. \eqref{Mmoments} is replaced by combinations of spherical Bessel functions. For example, for the zero-energy transition $2s\rightarrow2p$ Eqs. \eqref{LEMWightman} and \eqref{MEMWightman} read
\begin{align}
\mathcal{L}^\textsc{em}=&\,e^2 \frac{3a_0^2}{2\pi^2}\!\!\int_0^\infty\!\!\!\!\text{d}|\bm k||\bm k|^3 \frac{(a_0^2 |\bm k|^2-1)^2}{\left(a_0^2 |\bm k|^2+1\right)^8}\text{Re}\left[\mathcal{T}_0(|\bm k|,T)\right], \label{LEM}\\
\mathcal{M}^\textsc{em}=&-e^2\frac{3a_0^2}{2\pi^2}\cos\vartheta\notag\\
   &\times\int_0^\infty\text{d}|\bm k|\,|\bm k|^3\frac{(a_0^2|\bm k|^2-1)^2}{\left(a_0^2 |\bm k|^2+1\right)^8}\mathcal{T}_{\Delta t}(|\bm k|,T)\notag\\
   &\times\left[j_{0}(|\bm k||\bm x_\textsc{a}-\bm x_\textsc{b}|)+j_{2}(|\bm k||\bm x_\textsc{a}-\bm x_\textsc{b}|)\right], \label{MEM}
\end{align}
where $\vartheta$ is the angle of the axis of symmetry of atom B's $2p$ orbital with respect to atom A's orbital.

Note that, despite the fact that the hypergeometric function appearing in the scalar nonlocal term (see Eq. \eqref{Mmoments}) has been substituted by a combination of spherical Bessel functions, this combination can still be upper-bounded by $1$ (and the same occurs in the gapped case studied in Ref. \cite{Pozas-Kerstjens2016}). This means that also in this case the magnitude of the nonlocal term $|\mathcal{M}^\textsc{em}|$ is upper-bounded by the local term $\mathcal{L}^\textsc{em}$, which means that no entanglement can be harvested from the electromagnetic vacuum with degenerate atomic probes if their radial functions were compactly supported. This argument contains, as a special case, that studied numerically in Refs. \cite{Braun2002,Braun2005} where the atoms were assumed to be pointlike.

One must however note that the atomic wave functions of an electron in a hydrogenlike atom do not have  compact support. Instead, the radial wave functions decay exponentially with the distance to the atomic center of mass. For this reason, one may be tempted to argue that the atoms can never be placed in spacelike-separated regions due to the always-existent overlap of their atomic wave functions, which will make the imaginary part of the Wightman function contribute, albeit suppressed by a factor of the overlap between the wave functions. Nevertheless, for the implementation proposed in Ref. \cite{Braun2002} with two quantum dots separated by a distance of $d=10\,\text{nm}\approx 190\,a_0$ (where $a_0$ is the Bohr radius), the overlap between the wave functions is on the order of $\int\text{d}|\bm x| |\bm x|^2 \psi^\textsc{a}(|\bm x|)\psi^\textsc{b}(|\bm x|)\approx e^{-190}\approx 10^{-83}$, which is definitely negligible as compared with the entanglement that gapped atoms could harvest at those distance scales (for a detailed study on how the non-compact support cannot be responsible for entanglement harvesting, check section IV.C of Ref. \cite{Pozas-Kerstjens2016}). In the examples of Ref. \cite{Pozas-Kerstjens2016} the atoms were declared effectively spacelike when separated by $10^4$ Bohr radii and their interaction (with Gaussian switching) was short enough so that $10^4 a_0 /c$ was more than 9 times the timescale of duration of the interaction. In that example, the overlap between the wave functions of the two atoms was of the order of $10^{-4343}$, which is effectively 0 for all practical purposes. Since the harvesting of entanglement due to the atomic wave functions overlap is negligible, our results carry over to the light-atom interaction.

\section{Instantaneous switchings} \label{sec:deltas}

Finally, let us explore the case in which gapped detectors interact for an infinitesimal amount of time with the field but with an infinite strength. This case is relevant due to its similarities with a gapless detector case: In the case of a delta switching, during the time of interaction the free dynamics of the detectors is effectively halted (roughly speaking the free Hamiltonian becomes negligible with respect to the delta-strength of the interaction Hamiltonian). This interaction is modeled by Dirac delta switching functions
\begin{equation}
\mathcal{X}_\nu(t)=\eta \delta(t-t_\nu),
\label{deltaswitch}
\end{equation}
where $\eta$ is a constant with dimensions of time. This switching will allow us to obtain analytical closed-form expressions even for $\Omega\neq 0$.

For the switching function specified by Eq. \eqref{deltaswitch} the local and nonlocal terms Eqs. \eqref{L} and \eqref{M} read
\begin{align}
\mathcal{L}=&\ac^2\eta^2\int\text{d}^n\bm k \frac{|\tilde{S}(\bm{k})|^2}{2|\bm{k}|}\left|\int_{-\infty}^{\infty}\!\!\!\!\text{d}t\,\delta(t)e^{\ii(|\bm{k}|+\Omega)t}\right|^2\notag\\
=&\ac^2\eta^2\int\text{d}^n\bm k \frac{|\tilde{S}(\bm{k})|^2}{2|\bm{k}|},\\
\mathcal{M}=&-\ac^2\eta^2\int\text{d}^n\bm k\,\frac{|\tilde{S}(\bm{k})|^2}{2|\bm{k}|} e^{\ii\bm{k}\cdot(\bm{x}_\textsc{a}-\bm{x}_\textsc{b})} \notag\\
&\int_{-\infty}^{\infty}\!\!\!\!\text{d}t_1\int_{-\infty}^{t_1}\!\!\!\!\text{d}t_2\,e^{-\ii|\bm{k}|(t_1-t_2)}e^{\ii\Omega(t_1+t_2)}\notag\\
&\times\left[\delta\left(t_1-t_\textsc{a}\right)\delta\left(t_2-t_\textsc{b}\right)+\delta\left(t_1-t_\textsc{b}\right)\delta\left(t_2-t_\textsc{a}\right)\right].
\label{Msimul}
\end{align}

In the case of non-simultaneous switchings $t_\textsc{a}\neq t_\textsc{b}$, the argument in Sec. \ref{sec:nooverlap} used for evaluating the time integrals of the nonlocal term \eqref{Msimul} applies: if detector B is switched on after detector A, the first summand evaluates to zero while in the second the integrals denest. Integration over the time variables then leads to the expression
\begin{equation}
    \mathcal{M}=\!-\ac^2\eta^2\!\!\int\!\!\text{d}^n\bm k\,\frac{|\tilde{S}(\bm{k})|^2}{2|\bm{k}|} e^{\ii\bm{k}\cdot(\bm{x}_\textsc{a}-\bm{x}_\textsc{b})}e^{-\ii|\bm{k}|(t_\textsc{b}-t_\textsc{a})}e^{\ii\Omega(t_\textsc{b}+t_\textsc{a})}.
\label{eq:M-diracdeltas}
\end{equation}

The magnitude of this expression satisfies
\begin{align}
    |\mathcal{M}|\!=&\left|\ac^2\eta^2\!\!\int\!\!\text{d}^n\bm k\,\frac{|\tilde{S}(\bm{k})|^2}{2|\bm{k}|} e^{\ii\bm{k}\cdot(\bm{x}_\textsc{a}-\bm{x}_\textsc{b})}e^{-\ii|\bm{k}|(t_\textsc{b}-t_\textsc{a})}e^{\ii\Omega(t_\textsc{b}+t_\textsc{a})}\right|\notag\\
    \leq& \ac^2\eta^2\!\!\int\!\!\text{d}^n\bm k\,\frac{|\tilde{S}(\bm{k})|^2}{2|\bm{k}|} \left|e^{\ii\bm{k}\cdot(\bm{x}_\textsc{a}-\bm{x}_\textsc{b})}e^{-\ii|\bm{k}|(t_\textsc{b}-t_\textsc{a})}e^{\ii\Omega(t_\textsc{b}+t_\textsc{a})}\right|\notag\\
    =&\ac^2\eta^2\int\text{d}^n\bm k\,\frac{|\tilde{S}(\bm{k})|^2}{2|\bm{k}|}=\mathcal{L}
\end{align}
so, again in this case $\mathcal{N}^{(2)}=0$, regardless of the specific shape of the detectors, their relative distance, and additionally now the energy gap. 

When the individual interactions of the detectors with the field coincide, i.e. $\Delta t=0$, 
Eq. \eqref{eq:M-diracdeltas} becomes mathematically ambiguous, 
due to the argument of a Dirac delta coinciding with 
a limit of the integral. For sufficiently 
symmetric regularizations of the Dirac deltas, 
we however have
\begin{align}
& 2\int_{-\infty}^\infty\!\!\!\!\text{d}t_1\int_{-\infty}^{t_1}\!\!\!\!\text{d}t_2\,e^{-\ii|\bm k|(t_1-t_2)}e^{\ii\Omega(t_1+t_2)}
\delta(t_1 - t_\textsc{a})\delta(t_2 - t_\textsc{a})
\notag
\\[1ex]
& \hspace{1ex} = e^{2\ii\Omega t_\textsc{a}}, 
\label{eq:T-amb-interpretation}
\end{align}
and we give in appendix \ref{app:instant} 
two examples of such regularizations. With the interpretation \eqref{eq:T-amb-interpretation}, 
the nonlocal term $\mathcal{M}$ becomes 
\begin{equation}
    \mathcal{M}=-\ac^2 e^{2\ii\Omega t_\textsc{a}} \int\text{d}^n\bm k\,\frac{|\tilde{S}(\bm{k})|^2}{2|\bm{k}|} e^{\ii\bm{k}\cdot(\bm{x}_\textsc{a}-\bm{x}_\textsc{b})}.
\end{equation}

Again, the magnitude of this term is bounded from above by the local term $\mathcal{L}$, so $\mathcal{N}^{(2)}=0$ and entanglement harvesting is not possible in the limit when the switching becomes very short and intense, regardless of the shape or size of the probes, their relative distance or, in this specific case, the size of the gap between the energy levels of the detectors.

\section{Summary and discussion}\label{sec:summary}

In the context of entanglement harvesting \cite{Valentini1991,Reznik2005,Pozas-Kerstjens2015} and creation of entanglement by interaction with a common heat bath \cite{Braun2002,Braun2005}, we have studied whether degenerate identical two-level quantum systems coupling linearly with the vacuum state of a scalar field in flat spacetime are capable of harvesting the entanglement present in spacelike separated regions of the field. We have established several results within leading order in perturbation theory.

First, we have proved that if the time intervals of interaction between each individual detector and the field have no overlap the detectors can never become entangled through their interaction with the field. This result is independent of the shape or size of the detectors (which can be even not compactly supported in a finite region), the duration of the interaction or the separation between the probes (timelike, lightlike or spacelike).

Second, under the additional assumption of spherical symmetry of the detectors' smearing functions we have shown that, although the detectors can harvest timelike entanglement, for arbitrary spacelike separations entanglement harvesting is impossible in any situation where the time of interaction with the field is finite.

Third, we have shown that considering realistic light-matter interactions, and in particular the interaction of fully featured hydrogenlike atoms interacting with the electromagnetic field, the same phenomenology occurs: as the gap between the atomic levels is scaled down to 0 the gapless detectors become unable to harvest spacelike entanglement from the field, and only when the time intervals of the individual atomic interactions with the field overlap can the atoms have a chance of harvesting timelike and lightlike entanglement.

Finally, we have also shown that detectors coupled to the field through a delta-like coupling (short and intense coupling strength) are also completely unable to become entangled through their interaction with the field in timelike, spacelike or lightlike regimes at leading order in perturbation theory, regardless of their spatial smearing and, in this case, even if they have a finite energy gap. This should not be surprising since the delta coupling resembles a case where the detectors' internal dynamics are frozen during the time of interaction, as is the case of zero-gap detectors.

Therefore, we attribute the inability of gapless detectors to harvest entanglement to the fact that, as shown in previous studies \cite{Pozas-Kerstjens2015,Pozas-Kerstjens2016}, the energy gap has a protective role that shields from local noise allowing for nonlocal excitations that entangle the detectors. In the absence of a gap between the energy levels, even the smoothest switchings (those that create the smallest amount of local noise) break the entanglement between the detectors.

As a last comment, these results may also shed some light on studies in the context of creation of entanglement via interaction with a common heat bath through dipolar couplings \cite{Braun2002,Braun2005}. In these studies, the author saw numerically that only when one probe is deep inside the light cone of the other (they are in timelike separation) entanglement can be extracted from the bath to the (gapless) detectors. 

\begin{acknowledgements}
The authors thank Daniel Braun for the interesting conversations that motivated this work. 

The work of \mbox{A. P.-K.} is supported by Fundaci\'on Obra \mbox{Social} \mbox{``la Caixa''}, Spanish MINECO (QIBEQI \mbox{FIS2016-80773-P} and Severo Ochoa SEV-2015-0522), Fundaci\'o Privada Cellex and the Generalitat de Catalunya (SGR875 and CERCA Program). 

The work of \mbox{J. L.} is supported in part by the Science and Technology Facilities Council (Theory Consolidated Grant ST/J000388/1). 
\mbox{J. L.} thanks the Institute for Quantum Computing at the University of Waterloo for hospitality. 

The work of E. M.-M. is supported by the National Sciences and Engineering Research Council of Canada through the Discovery program. E. M.-M. also thankfully acknowledges the funding of his Ontario Early Research Award.
\end{acknowledgements}

\appendix

\section{Integration over angular variables of the nonlocal term}\label{app:nonlocalangleintegral}
In this appendix we perform the integrations in the generalized solid angle variables of the vector $\bm k$ that appear in the nonlocal term $\mathcal{M}$ 
in Eq. \eqref{Mangles}, namely
\begin{equation}
    \int\text{d}\Omega_{n-1}\,e^{\ii\bm k\cdot(\bm x_\textsc{a}-\bm x_\textsc{b})},
\end{equation}
to compare the result to the corresponding integrals in the local term $\mathcal{L}$, which evaluate to the area of the $(n-1)$-sphere, i.e.,
\begin{equation}
    \int\text{d}\Omega_{n-1}=A_{n-1}=\frac{2\pi^{\frac{n}{2}}}{\Gamma(n/2)}.
    \label{sphere}
\end{equation}


In $n$ dimensions there are $n-1$ angular variables, one of which (the polar angle $\phi_{n-1}$) has the range $[0,2\pi)$ and the rest (the azimuthal angles $\phi_1,\dots\phi_{n-2}$) have range $[0,\pi]$. The solid angle element is therefore
\begin{align}
    \text{d}\Omega_{n-1}=&\sin^{n-2}(\phi_{1})\sin^{n-3}(\phi_{2})\dots\sin(\phi_{n-2})\notag\\
    &\times\text{d}\phi_1\text{d}\phi_2\dots\text{d}\phi_{n-1}.
\end{align}

Let us then begin with the particularly simple case of $n=2$ for illustration. Choosing the $x$ axis of the integration frame to align with $\bm x_\textsc{a}-\bm x_\textsc{b}$, the integral easily evaluates to (see 10.9.4 and 10.16.9 in Ref. \cite{NIST:DLMF})
\begin{align}
\int_0^{2\pi}\text{d}\phi_1\,e^{\ii|\bm k||\bm x_\textsc{a}-\bm x_\textsc{b}|\cos\phi_1}=&2\pi J_0(|\bm k||\bm x_\textsc{a}-\bm x_\textsc{b}|)\notag\\
=&2\pi\,{}_0F_1\left(\!1;-\frac{(|\bm k||\bm x_\textsc{a}-\bm x_\textsc{b}|)^2}{4}\right)\!,
\label{twodimangular}
\end{align}
where ${}_0F_1$ is the confluent hypergeometric limit function.

In fact, the general case is not too difficult to compute either. In $n$ spatial dimensions, one can choose to place one of the axes of the integration frame aligned with \mbox{$\bm x_\textsc{a}-\bm x_\textsc{b}$}, which simplifies the scalar product in the exponential to, for instance, \mbox{$|\bm k||\bm x_\textsc{a}-\bm x_\textsc{b}|\cos(\phi_{1})$}. With this choice, the integrals evaluate to
\begin{align}
    \int\text{d}\Omega_{n-1}\,&e^{\ii\bm k\cdot(\bm x_\textsc{a}-\bm x_\textsc{b})}=2\pi \prod_{m=2}^{n-2}\frac{\sqrt{\pi } \Gamma \left(\frac{m}{2}\right)}{\Gamma \left(\frac{m+1}{2}\right)}\notag\\
    &\times\int_0^\pi\text{d}\phi_{1}\,\sin^{n-2}(\phi_{1})e^{\ii|\bm k||\bm x_\textsc{a}-\bm x_\textsc{b}|\cos(\phi_{1})}\notag\\
    =&2\pi \left(\prod_{m=2}^{n-2}\frac{\sqrt{\pi } \Gamma \left(\frac{m}{2}\right)}{\Gamma \left(\frac{m+1}{2}\right)}\right)\sqrt{\pi } \frac{\Gamma \left(\frac{n-1}{2}\right)}{\Gamma \left(\frac{n}{2}\right)}\notag\\
    &\times {}_0 F_1\left(\frac{n}{2};-\frac{(|\bm k||\bm x_\textsc{a}-\bm x_\textsc{b}|)^2}{4}\right)\notag\\
    =&\frac{2\pi ^{\frac{n}{2}}}{\Gamma(n/2)}{}_0F_1\left(\frac{n}{2};-\frac{(|\bm k||\bm x_\textsc{a}-\bm x_\textsc{b}|)^2}{4}\right),
    \label{angularintegrals}
\end{align}
using again 10.9.4 and 10.16.9 in Ref. \cite{NIST:DLMF}, 
and noting that 
\begin{equation}
\prod_{i=k}^{l}f_i\coloneqq 1 \qquad\text{for}\,\,l<k,
    \label{product}
\end{equation}
and
\begin{equation}
    \prod_{m=2}^{n-2}\frac{\sqrt{\pi } \Gamma \left(\frac{m}{2}\right)}{\Gamma \left(\frac{m+1}{2}\right)}
    =\begin{cases}
    1 & n \leq3\\
    \frac{\pi ^{\frac{n-3}{2}}}{\Gamma \left(\frac{n-1}{2}\right)} & n\geq 3
    \end{cases}.
\end{equation}



\section{Time integrals in the overlapping case}\label{app:time}

In this appendix we examine the time integrals 
in the local term Eq. \eqref{L}, 
given by  
\begin{align}
\mathcal{T}_\mathcal{L}(|\bm k|,T)
&=
\int_{-\infty}^\infty\!\!\!\!\text{d}t_1\int_{-\infty}^\infty\!\!\!\!\text{d}t_2\,e^{-\ii|\bm{k}|(t_1-t_2)}\mathcal{X}\left(t_1\right)\mathcal{X}\left(t_2\right)
\notag\\
&=2\pi|\tilde{\mathcal{X}}(|\bm k|)|^2, 
\end{align}
where the tilde denotes the Fourier transform in the notation of Eq. \eqref{Fourier}. 
We show that 
\begin{align}
\mathcal{T}_\mathcal{L}(|\bm k|,T)
=
\text{Re}\left[\mathcal{T}_0(|\bm k|, T)\right],
\end{align}
where $\mathcal{T}_{\Delta t}(|\bm k|, T)$ is given by Eq. \eqref{TtB}.

To begin with, we see that for $\Delta t=0$ the two summands of Eq. \eqref{TtB} coincide, leading to
\begin{align}
\mathcal{T}_0(|\bm k|,T)=&2\int_{-\infty}^\infty\!\!\!\!\text{d}t_1\int_{-\infty}^\infty\!\!\!\!\text{d}t_2\,e^{-\ii|\bm{k}|(t_1-t_2)}\notag\\
&\times \mathcal{X}\left(t_1\right)\mathcal{X}\left(t_2\right)\theta(t_1-t_2),
\end{align}
where $\theta(x)$ is the Heaviside step function. Using the identity $1=\theta(x)+\theta(-x)$ and performing the change of variables $t_1\leftrightarrow t_2$ in the second summand the result follows:
\begin{align}
\mathcal{T}_{\mathcal{L}}(|\bm k|,T)&=\int_{-\infty}^\infty\!\!\!\!\text{d}t_1\int_{-\infty}^\infty\!\!\!\!\text{d}t_2\,e^{-\ii|\bm{k}|(t_1-t_2)}\notag\\
&\qquad\times\mathcal{X}\left(t_1\right)\mathcal{X}\left(t_2\right)\left[\theta(t_1-t_2)+\theta(t_2-t_1)\right]\notag\\
&=\int_{-\infty}^\infty\!\!\!\!\text{d}t_1\int_{-\infty}^\infty\!\!\!\!\text{d}t_2\,\mathcal{X}\left(t_1\right)\mathcal{X}\left(t_2\right)\notag\\
&\qquad\times\theta(t_1-t_2)\left(e^{-\ii|\bm{k}|(t_1-t_2)}+e^{\ii|\bm{k}|(t_1-t_2)}\right)\notag\\
&=\int_{-\infty}^\infty\!\!\!\!\text{d}t_1\int_{-\infty}^\infty\!\!\!\!\text{d}t_2\,\mathcal{X}\left(t_1\right)\mathcal{X}\left(t_2\right)\notag\\
&\qquad\times\theta(t_1-t_2)2\text{Re}\left(e^{-\ii|\bm{k}|(t_1-t_2)}\right)\notag\\
&=\text{Re}\left[\mathcal{T}_0(|\bm k|,T)\right].
\label{timelocal}
\end{align}

\section{Evaluation of $\Realpart\mathcal{T}_{\Delta t}$}\label{app:TtB}

In this appendix we show that Eq. \eqref{TtB} leads to Eq. \eqref{TtBsimple}.

Starting from Eq. \eqref{TtB} and changing variables by \mbox{$t_1=t_2+s$} gives 
\begin{align}
    \mathcal{T}_{\Delta t}&(|\bm k|,T)=\int_{-\infty}^{\infty}\!\!\!\!\text{d}t_2\int_{0}^{\infty}\!\!\!\!\text{d}s\,e^{-\ii|\bm k|s}\notag\\
    &\times\left[\mathcal{X}\left(t_2+s\right)\mathcal{X}\left(t_2-\Delta t\right)+\mathcal{X}\left(t_2+s-\Delta t\right)\mathcal{X}\left(t_2\right)\right].
\end{align}

Changing variables in the first summand by $\mu=t_2+s$
and renaming $\mu=t_2$ in the second summand, we obtain 
\begin{align}
    \mathcal{T}_{\Delta t}(|\bm k|,T)=&\int_{-\infty}^{\infty}\!\!\!\!\text{d}\mu \, \mathcal{X}
    \left(\mu\right) \int_{0}^{\infty}\!\!\!\!\text{d}s\,e^{-\ii|\bm k|s}\notag\\
    &\times\left[\mathcal{X}\left(\mu-s-\Delta t\right)+\mathcal{X}\left(\mu+s-\Delta t\right)\right].
\end{align}

Taking the real part gives 
\begin{align}
&\text{Re}\left[\mathcal{T}_{\Delta t}(|\bm k|,T)\right]
= 
\int_{-\infty}^{\infty}\!\!\!\!\text{d}\mu \, \mathcal{X} \left(\mu\right) 
\notag\\
&\hspace{2ex}
\times 
\int_{0}^{\infty}\!\!\!\!\text{d}s \cos(|\bm k|s)
\left[\mathcal{X}\left(\mu-s-\Delta t\right)+\mathcal{X}\left(\mu+s-\Delta t\right)\right]
\notag\\
&
= 
\frac12 \int_{-\infty}^{\infty}\!\!\!\!\text{d}\mu \, \mathcal{X} \left(\mu\right) 
\notag\\
&\hspace{2ex}
\times 
\int_{-\infty}^{\infty}\!\!\!\!\text{d}s \, e^{\ii|\bm k|s}
\left[\mathcal{X}\left(\mu-s-\Delta t\right)+\mathcal{X}\left(\mu+s-\Delta t\right)\right]
\notag\\
&=
\frac{\sqrt{2\pi}}{2}
\int_{-\infty}^{\infty}\!\!\!\!\!\!\text{d}\mu\,\mathcal{X}\left(\mu\right)\!\!
\notag\\
&\hspace{2ex}
\times
\left[e^{\ii|\bm k|(\mu-\Delta t)}[\tilde{\mathcal{X}}\left(|\bm k|\right)]^*
+e^{\ii|\bm k|(-\mu+\Delta t)}\tilde{\mathcal{X}}\left(|\bm k|\right)\right]
\notag\\
&= 
\pi\left[|\tilde{\mathcal{X}}(|\bm k|)|^2e^{-\ii|\bm k|\Delta t}
+|\tilde{\mathcal{X}}(|\bm k|)|^2e^{\ii|\bm k|\Delta t}\right]
\notag\\
&= 2\pi|\tilde{\mathcal{X}}(|\bm k|)|^2\cos(|\bm k|\Delta t),
\end{align}
where the second equality uses the evenness of 
$\mathcal{X}\left(\mu-s-\Delta t\right)+\mathcal{X}\left(\mu+s-\Delta t\right)$ in~$s$.

\section{Regularizations of instantaneous switching}\label{app:instant}

In this appendix we present two regularizations of the Dirac delta that are `kink' limits of switchings largely employed in past literature \cite{Pozas-Kerstjens2015} that 
lead to \eqref{eq:T-amb-interpretation}. 
For notational simplicity, we set $t_\textsc{a}=0$ and consider the formal expression 
\begin{align}
\mathcal{T}_0(|\bm k|) 
= 
2\int_{-\infty}^\infty\!\!\!\!\text{d}t_1\int_{-\infty}^{t_1}\!\!\!\!\text{d}t_2\,e^{-\ii|\bm k|(t_1-t_2)}e^{\ii\Omega(t_1+t_2)}
\delta(t_1)\delta(t_2), 
\end{align}
showing that each of the regularizations gives for $\mathcal{T}_0(|\bm k|)$ the value unity. 

\subsection{Top-hat regularization}

First, we regard the Dirac delta as a limit of the top-hat function, 
\begin{equation}
    \delta(t)=\lim_{\epsilon\to 0^+}\frac{1}{\epsilon}\begin{cases}
    1 & \text{if}\,\,t\in\left[-\frac{\epsilon}{2},\frac{\epsilon}{2}\right]\\
    0 & \text{otherwise}
    \end{cases}.
\end{equation}
Then
\begin{align}
    \mathcal{T}_0(|\bm k|)=&2\int_{-\infty}^\infty\!\!\!\!\text{d}t_1\int_{-\infty}^{t_1}\!\!\!\!\text{d}t_2\,e^{-\ii|\bm k|(t_1-t_2)}e^{\ii\Omega(t_1+t_2)}\delta(t_1)\delta(t_2)\notag\\
    =&2\!\!\lim_{\epsilon\to 0^+}\lim_{\epsilon^\prime\to 0^+}\frac{1}{\epsilon\epsilon^\prime}\!\int_{-\epsilon/2}^{\epsilon/2}\!\!\!\!\!\!\!\!\!\!\text{d}t_1\int_{-\epsilon^\prime/2}^{t_1}\!\!\!\!\!\!\!\!\!\!\text{d}t_2\,e^{-\ii|\bm k|(t_1-t_2)}e^{\ii\Omega(t_1+t_2)}\notag\\
    =&2\lim_{\epsilon\to 0^+}\lim_{\epsilon^\prime\to 0^+}\frac{\ii e^{-\frac{1}{2} \ii\epsilon^\prime (|\bm k|+\Omega )}}{\epsilon\epsilon^\prime (|\bm k|+\Omega )}\notag\\
    &\times\int_{-\epsilon/2}^{\epsilon/2}\!\!\!\!\text{d}t_1\,e^{-\ii (|\bm k|-\Omega )t_1} \left(1-e^{\frac{1}{2} \ii (|\bm k|+\Omega ) (2 t_1+\epsilon^\prime)}\right)\notag\\
    =&\lim_{\epsilon\to 0^+}\lim_{\epsilon^\prime\to 0^+}\frac{2\ii}{\epsilon\epsilon^\prime}\left[\frac{2 e^{-\frac{1}{2} \ii \epsilon^\prime (|\bm k|+\Omega )} \sin \left[\frac{1}{2} \epsilon  (|\bm k|-\Omega )\right]}{|\bm k|^2-\Omega ^2}\right.\notag\\
    &\left.-\frac{\sin (\Omega  \epsilon )}{|\bm k| \Omega +\Omega ^2}\right.\Bigg]\notag\\
    =&\lim_{\epsilon\to 0^+}\lim_{\epsilon^\prime\to 0^+}\frac{2\ii}{\epsilon\epsilon^\prime}\frac{(-\ii\epsilon\epsilon^\prime)}{2}=\,1.
\end{align}

\phantom{Spaceforpresentation}

\subsection{Gaussian regularization}

Second, we regard the Dirac delta as a limit of the Gaussian function, 
\begin{equation}
\delta(t)=\lim_{\epsilon\to 0^+}\frac{1}{2\epsilon\sqrt{\pi}}e^{-\frac{t^2}{4\epsilon^2}}. 
\end{equation}
Then
\begin{align}
    \mathcal{T}_0(|\bm k|)=&\lim_{\epsilon\to 0^+}\lim_{\epsilon^\prime\to 0^+}\frac{1}{2\pi\epsilon\epsilon^\prime}\int_{-\infty}^{\infty}\!\!\!\!\text{d}t_1\,e^{-\ii(|\bm k|-\Omega)t_1}e^{-\frac{t_1^2}{4\epsilon^2}}\notag\\
    &\times \int_{-\infty}^{t_1}\!\!\!\!\text{d}t_2\,e^{\ii(|\bm k|+\Omega)t_2}e^{-\frac{t_2^2}{4{\epsilon^\prime}^2}}\notag\\
    =&\lim_{\epsilon\to 0^+}\lim_{\epsilon^\prime\to 0^+}\frac{e^{-{\epsilon^\prime}^2(|\bm k|+\Omega)^2}}{2\sqrt{\pi}\epsilon}\int_{-\infty}^{\infty}\!\!\!\!\text{d}t_1\,e^{-  \ii(|\bm k|-\Omega)t_1}\notag\\
    &\times e^{-\frac{t_1^2}{4\epsilon^2}}\left[1+\text{erf}\left(\frac{t_1}{2\epsilon^\prime}-\ii\epsilon^\prime(|\bm k|+\Omega)\right)\right].
\end{align}

This is exactly Eq. (A2) in appendix A of Ref. \cite{Pozas-Kerstjens2015}. As shown there, the remaining integral has a closed-form expression, which yields
\begin{align}
    \mathcal{T}_0(|\bm k|)\!=&\!\lim_{\epsilon\to 0^+}\lim_{\epsilon^\prime\to 0^+}e^{-\epsilon^2(|\bm k|+\Omega)^2}e^{-{\epsilon^\prime}^2(|\bm k|-\Omega)^2}\notag\\
    &\times\!\left[1\!+\!\text{erf}\left(\ii\frac{\epsilon(|\bm k|-\Omega)+\epsilon^\prime(|\bm k|+\Omega)}{\sqrt{2}}\right)\right]=\,1.
\end{align}

\phantom{Spaceforpresentation}

\bibliography{bibliography}

\end{document}